\documentclass[preprint,prl,aps]{revtex4}

\begin{document}
\newcommand{\be}{\begin{equation}}
\newcommand{\ee}{\end{equation}}
\newcommand{\ben}{\begin{eqnarray}}
\newcommand{\een}{\end{eqnarray}}
\newcommand{\n}{\nonumber  }
\newcommand{\nn}{\nonumber \\ }
\newcommand{\nd}{\noindent}
\newcommand{\p}{\partial}

\title{Direct Fisher inference of the quartic  oscillator's eigenvalues}

\author{S.P. Flego$^1$}
\author{A. Plastino$^{2}$}
\author{A.R. Plastino$^{3,\,4}$}

\affiliation{ $^{1}$Universidad Nacional de La Plata, Facultad de Ingenier\'{\i}a, \'Area Departamental de Ciencias B\'asicas,
 1900 La Plata, Argentina \\  $^{2}$Universidad Nacional de La Plata, Instituto
de F\'{\i}sica (IFLP-CCT-CONICET), C.C. 727, 1900 La Plata, Argentina \\
$^{3}$CREG-Universidad Nacional de La Plata-CONICET, C.C. 727, 1900 La Plata, Argentina\\
$^{4}$Instituto Carlos I de Fisica Teorica y Computacional and
Departamento de Fisica Atomica, Molecular y Nuclear, Universidad
de Granada, Granada, Spain}
%

\begin{abstract}
\nd  It is well known that a suggestive connection  links
Schr\"odinger's equation (SE) and the information-optimizing
principle based on Fisher's information measure (FIM). It has been
shown that this entails the existence of a Legendre transform
structure underlying the SE. Such a structure leads to a first
order partial differential equation (PDE) for the SE's eigenvalues from which a
complete solution for them can be obtained. As an application we
deal with the quantum theory of anharmonic oscillators, a
long-standing problem that has received intense attention
motivated by problems in quantum field  theory and molecular
physics. By appeal to the Cramer Rao bound we are able to Fisher-infer
the particular PDE-solution that yields the eigenvalues without
explicitly solving Schr\"odinger's equation. Remarkably enough,
and in contrast with standard variational approaches, our present
procedure does not involve free fitting parameters.

\vspace{2.cm}


\nd PACS: 05.45+b, 05.30-d
\end{abstract}


\maketitle

\section{1. Introduction}

\nd It is well-known that a strong link exists between  Fisher'
information measure (FIM) $I$ \cite{frieden2} and Schr\"odinger
wave equation (SE)
\cite{pla7,flego,reginatto,HF-TV-RR,ArX1Univ,ArX2Alpha,ArX3}. In a
nutshell, this connection is based upon the fact that a
constrained Fisher-minimization leads to a SE-like equation
\cite{frieden2,pla7,flego,reginatto,HF-TV-RR,ArX1Univ,ArX2Alpha,ArX3}.
In turn, this implies the existence of intriguing relationships
between various  SE-facets,    and  Jaynes's maximum entropy
principle. In particular, basic SE-consequences such as the
Hellmann-Feynman and the Virial theorems can be re-interpreted in
terms of a special kind of reciprocity relations between relevant
physical quantities, similar to the ones exhibited by the
thermodynamics' formalism via its Legendre-invariance property
\cite{ArX1Univ,HF-TV-RR}. This fact demonstrates that a
Legendre-transform structure underlies the non-relativistic
Schr\"odinger equation. As a consequence, the possible
energy-eigenvalues are now seen to be constrained by such
structure in a rather unsuspected way
\cite{HF-TV-RR,ArX1Univ,ArX2Alpha,ArX3}, a fact that allows one to
obtain a first-order differential equation, unrelated to
Schroedinger's equation \cite{ArX2Alpha,ArX3},  that energy
eigenvalues must necessarily satisfy. The predictive power of this
new equation will be explored here.

\nd We will apply our formalism here to the quantum anharmonic oscillator,
which is the paradigmatic testing-ground for new approaches to Schroedinger
eigenvalue equation. Besides their intrinsic conceptual and mathematical interest,
anharmonic oscillators have received considerable attention over the years due
to their practical relevance in connection with several areas of physics, such as
quantum field theory and molecular physics, among others. In this kind of systems,
the most intense focus has been traditionally concentrated upon the quartic
oscillator. General accounts containing illuminating  references
on this problem may be found, for instance, in
\cite{Hioe-75,Baner-I}. Note that a perturbation series solution
to this problem in powers of the anharmonicity-parameter $\lambda$
is divergent-asymptotic for all $\lambda>0$ \cite{Bender-69}.
Specifically,  we will apply our procedure to treat the quartic
anharmonic oscillator. By appeal to the Cramer Rao bound we obtain
the particular solution that leads to the system's eigenvalues
without need of  explicitly solving Schr\"odinger's equation. More
importantly, we do not need to appeal to any arbitrary or
empirical parameter, as is common practice in other treatments
\cite{Banerjee}.

\section{2. Basic ideas}

\nd Let $x$ be a stochastic variable  and $f(x,\theta)$ the
probability density function (PDF) for this variable, which
depends on the parameter $\theta$. If an observer were to make a
measurement of $x$ and had to best infer  $\theta$ from such
measurement, calling the resulting estimate $\tilde \theta=\tilde
\theta(x)$, one might ask how well could $\theta$ be determined.
Estimation theory~\cite{frieden2} tells us that the {\it best
possible estimator} $\tilde \theta( x)$, after a very large number
of $x$-samples is examined, suffers a mean-square error $\Delta x$ from
$\theta$ obeying the rule $I~(\Delta x)^2=1$, where $I$ is an information
quantifier called the Fisher information measure (FIM), a non
linear  functional of the PDF that reads
 \be \label{eq.1-1} I \,=\,\int ~dx ~f(x,\theta)
\left\{\frac{\partial ~ }{\partial
\theta}~\ln{[f(x,\theta)]}\right\}^2.\ee Any other estimator must
have a larger mean-square error (all estimators must be unbiased,
i.e., satisfy $ \langle \tilde \theta({\bf x}) \rangle=\,\theta
\label{unbias}$). Thus, FIM has a lower bound. No matter what the
parameter $\theta$ might be, $I$ has to obey \be \label{rao}
I\,(\Delta x)^2\,\ge \,1,\ee the  celebrated Cramer--Rao bound
\cite{frieden2}.

\nd In the case of physical Fisher applications the particular
instance of translational families merits special consideration.
These are mono-parametric distribution families of the form
$f(x,\theta)=f(x-\theta),$ known up to the shift parameter
$\theta$. All family members exhibit identical shape. For such
families we get \ben \label{eq.1-2}  I = \, \int f(x)
\left(\frac{\partial \ln{f(x)} }{\partial x} \right)^2 dx. \een
Focus attention now a system that is specified by a set of $M$
physical parameters $\mu_k$. We can write $\mu_k = \langle
A_{k}\rangle,$ with  $A_{k}= A_{k}(x).$ The set of
$\mu_{k}$-values is to be regarded as our prior knowledge. It
represents our available empirical information. Let the pertinent
probability distribution function (PDF) be $f(x)$. Then, \be
\label{eq.1-3} \langle A_{k}\rangle\,=\,\int ~dx ~A_{k}(x) ~f(x),
\hspace{0.5cm} k=1,\dots ,M. \ee In this context it can be shown
(see for example \cite{pla7,reginatto}) that the {\it physically
relevant} PDF $f(x)$  minimizes  FIM subject to
the prior conditions and the normalization condition.
Normalization entails $\int dx  f(x) = 1,$  and, consequently, our
Fisher-based extremization problem adopts the appearance \be
\label{eq.1-4}\delta \left( I - \alpha \int ~dx ~f(x) -
\sum_{k=1}^M~\lambda_k\int ~dx ~A_{k}(x)~f(x)\right) = ~0, \ee
where we have introduced the $(M+1)$ Lagrange multipliers
$\lambda_k$ ($\lambda_0=\alpha$). In Ref. \cite{pla7} on can find
the details of how to go  from (\ref{eq.1-4}) to a Schr\"odinger's
equation (SE) that yields the desired PDF in terms of the
amplitude $\psi(x)$ defined by $f(x)=\psi(x)^2$. This SE is of the
form
 \be \label{eq.1-5} \left[-~\frac{1}{2}~\frac{\partial^2 ~}{\partial x^2} +U(x)\right]\psi ~= ~ \frac{\alpha}{8}~ \psi,\hspace{1.3cm} U(x)
=~-\frac{1}{8}~\sum_{k=1}^{M}\,\lambda_{k}~ A_{k}(x),\ee
and can be formally interpreted as the (real) Schr\"odinger
equation (SE) for a particle of unit mass ($\hbar=1$) moving in the
effective, ``information-related pseudo-potential" $U(x)$ \cite{pla7}
 in which the normalization-Lagrange multiplier ($\alpha /8$) plays the role
of an energy eigenvalue. The  $\lambda_k$ are fixed, of course, by
recourse to the available prior information. In the case of
one-dimensional scenarios,   $\psi(x)$ is real \cite{richard} and
\ben \label{eq.1-7}  I =   \, \int  \psi^2
\left(\frac{\partial \ln{\psi^2} }{\partial x} \right)^2 dx\,=\, 4
\int  \left(\frac{\partial \psi }{\partial x} \right)^2 dx
 =\, -4 \int  \psi \frac{\partial^2 ~}{\partial x^2} \psi~dx
 \een
\nd so that using the SE (\ref{eq.1-5}) we obtain
\ben \label{eq.1-12} I=\,\alpha
 + \sum_{k=1}^M~\lambda_k\left\langle A_k\right\rangle. \een

\vskip 2mm  \nd   {\bf Legendre structure}

\vskip 2mm

\nd The connection between the variational solutions $f$  and
thermodynamics was established in Refs. \cite{pla7} and \cite{flego} in the guise of
reciprocity relations that express the Legendre-transform structure of  thermodynamics.
They constitute its essential formal ingredient \cite{deslog}
and were re-derived \`a la Fisher in \cite{pla7} by recasting (\ref{eq.1-12})
in a fashion that emphasizes the role of the relevant independent
variables,
\ben \label{eq.1-13a} I(\langle
A_1\rangle,\ldots,\langle A_M\rangle) \,=\,\alpha
  + \sum_{k=1}^M~\lambda_k \langle A_k\rangle. \een
 Obviously, the Legendre transform main goal is that of  changing the identity
of our relevant variables. As for  $I$ we have
\be \label{eq.1-13b}\alpha(\lambda_1,\ldots,\lambda_M)= I-
\sum_{k=1}^M~\lambda_k\left\langle A_k\right\rangle , \ee
so that we encounter the three reciprocity relations
(proved in \cite{pla7})
\be \label{RR-1} \frac{\partial \alpha}{\partial \lambda_{k}}= -
\langle A_k\rangle ~; \hspace{1.cm}  \frac{\partial I }{\partial
\langle A_k \rangle}\,=\,\lambda_k  ~ ;\hspace{1.cm}
\frac{\partial I}{\partial \lambda_{i}}=\sum_{k}^{M} \lambda_{k}
 \frac{\partial \langle A_k \rangle}{\partial \lambda_{i}},\ee
the last one being a generalized Fisher-Euler theorem.

\section{3. Fisher measure and quantum mechanical connection}

\nd The potential function $U(x)$ belongs to $\mathcal{L}_2$
 and thus admits of a series expansion in the basis
$x,\,x^2\,x^3,\,etc.$ \cite{greiner}. The $A_k(x)$ themselves
belong to $\mathcal{L}_2$ as well and can also  be series-expanded
in similar fashion. This enables us to base our future
considerations on the assumption that the a priori knowledge
refers to moments $x^k$ of the independent variable, i.e., $
\langle A_k \rangle~=~ \langle x^k \rangle $, and that one
possesses information about  $M$ of these moments
 $\langle x^k \rangle$. Our ``information" potential $U$
  then reads
\be \label{virial-5} U(x)=  -~ \frac{1}{8} \sum_k \,\lambda_k\,
x^k. \ee  {\it We will assume that the first $M$ terms of the
above series yield a satisfactory
 representation of} $~U(x)$. Consequently, the Lagrange multipliers are
 identified with  U(x)'s series-expansion's coefficients.

\nd In this Schr\"odinger-scenario the {\it virial theorem} states that \cite{HF-TV-RR}
 \ben \label{virial-6}
\left\langle \frac{\partial^2~}{\partial x^2}\right\rangle = -~
\left\langle x ~ \frac{\partial ~}{\partial x} U(x)\right\rangle ~
= ~ \frac{1}{8}~\sum_{k=1}^{M}\, k \,\lambda_{k}~\langle x^k
\rangle ~, \een and thus, from (\ref{eq.1-7}) and (\ref{virial-6})
a useful, virial-related expression for Fisher's information
measure can be arrived at \cite{HF-TV-RR}

\ben \label{virial-7} I\,
=\, -~ ~\sum_{k=1}^{M}\, \frac{k}{2} \,\lambda_{k}~\langle
x^{k}\rangle, \een  Also, substituting the above $I$-expression
into (\ref{eq.1-12}) and solving for $\alpha$, we obtain \ben
\label{virial-8} \alpha \,  =\,
-~\sum_{k=1}^{M}\,\left(1+\frac{k}{2}\right)\lambda_k ~\langle x^k
\rangle.  \een  $\alpha$ ($I$) is explicit function of the $M$
Lagrange multipliers - $U(x)$'s series-expansion coefficients
$\lambda_{k}$ (associated to  the physical parameters $\langle
x^{k}\rangle$).
  Eqs. (\ref{virial-7}) and (\ref{virial-8}) encode the
information provided by the virial theorem
\cite{ArX1Univ,HF-TV-RR}.

\subsection{Fisher-Schr\"oedinger Legendre structure}

\nd  Interestingly enough, the reciprocity relations (RR)
(\ref{RR-1}) can be re-derived on a strictly pure quantum
mechanical basis \cite{HF-TV-RR}, starting from
\begin{enumerate}
\item the quantum Virial theorem [which leads to Eqs.
(\ref{virial-7}) and (\ref{virial-8})] plus \item  information
provided by the quantum Hellmann-Feynman theorem. \end{enumerate}
 This fact indicates that a Legendre structure underlays  the one-dimensional
Schr\"oedinger equation \cite{HF-TV-RR}. Thus, with $\langle
A_k\rangle=\langle x^k\rangle$, our ``new"  reciprocity relations
are given by \be \label{RR-q} \frac{\partial \alpha}{\partial
\lambda_{k}}= - \langle x^k\rangle ~; \hspace{1.cm} \frac{\partial
I }{\partial \langle x^k \rangle}\,=\,\lambda_k  ~ ;\hspace{1.cm}
\frac{\partial I}{\partial \lambda_{i}}=\sum_{k}^{M} \lambda_{k}
 \frac{\partial \langle x^k \rangle}{\partial \lambda_{i}},\ee

\nd FIM expresses a relation between the independent variables or
control variables (the prior information) and $I$. Such
information is encoded into the functional form  $I=I(\langle x^1
\rangle, ... , \langle x^M \rangle  )$. For later convenience, we
will also denote such a relation or encoding as $\{I,\langle x^k
\rangle \}$. We see that the Legendre transform FIM-structure
involves eigenvalues of the ``information-Hamiltonian" and
Lagrange multipliers.
  Information is encoded in
 $I$ via these Lagrange multipliers, i.e., ${\alpha}={\alpha}(\lambda_1, ... \lambda_M),$ $
{\rm together \,\, with \,\, a  \,\, bijection} \,\,\,\{I,\langle
x^k \rangle \}  \hspace{0.6cm} \longleftrightarrow \hspace{0.6cm}
\{{\alpha}, \lambda_k \}. \label{RR-3}$

\subsection{Two scenarios}
\nd {\bf In a  $\left\{I, \langle {x}^{k}\rangle \right\}$ -
scenario}, the $\lambda_k$ are  functions dependent on the $\langle
 {x}^{k}\rangle$-values.
As  shown in \cite{ArX1Univ}, substituting the RR given by
(\ref{RR-q}) in (\ref{virial-7}) one is led to a {\it linear,
partial differential equations (PDE)} for $I$, \ben \label{gov-1}
\lambda_k \,=\, \frac{\partial I }{\partial \langle x^k \rangle}
\hspace{1.cm}\longrightarrow \hspace{1.cm}  I \,  =\,
-~\sum_{k=1}^{M}\,\frac{k}{2} ~\langle x^{k}\rangle ~
\frac{\partial I }{\partial \langle x^k \rangle}\,. \een and a
complete solution is given by \ben \label{gov-9} {I}(\langle {x}^1
\rangle, ... , \langle {x}^M \rangle ) =~
 \sum_{k=1}^{M}~C_k~ ~{ \left| \langle
{x}^{k}\rangle \right|^{- {2}/{k}}}~,\een
 where $C_k$ are positive real numbers (integration constants).
The $I$ - domain is ${\it D}_I=\left\{(\langle {x}^1 \rangle, ...
, \langle {x}^M \rangle) / \langle {x}^k \rangle~\in ~\Re_o
\right\}$.  Eq. (\ref{gov-9}) states that for $\langle {x}^k
\rangle >0$, $I$ is a monotonically decreasing function of $ \langle
x^k \rangle$, and as one expects from a ``good'' information
measure \cite{frieden2}, $I$ is a convex function. We may obtain
$\lambda_k$ from the reciprocity relations (\ref{RR-q}). For
$\langle x^k \rangle ~ > ~ 0 ~$ one gets,
\ben \label{prop-2-RR} \lambda_k ~=~ \frac{\partial I}{\partial \langle
x^k \rangle} ~=~-~ \frac{2}{k}~C_k~ ~{ \langle {x}^{k}\rangle^{-~
(2+k)/k}}~<~0~. \een and then, using (\ref{eq.1-12}), we obtain
the $\alpha$ - normalization Lagrange multiplier.

\nd The general solution for the $I$ - PDE does exist and its
uniqueness has been demonstrated via an analysis of the associated
Cauchy problem \cite{ArX1Univ}. Thus, Eq. (\ref{gov-9}) implies
what seems to be a kind of ``universal'' prescription, a  linear
PDE that any variationally (with constraints) obtained FIM must
necessarily comply with.

\vskip 3mm

\nd {\bf In the $\left\{\alpha, \lambda_k \right\}$ scenario},
the $\langle {x}^{k}\rangle$ are  functions that depend  on the
$\lambda_k$-values.
\nd As we showed in \cite{ArX2Alpha},
an analog {\it $\alpha$-PDE} exists.
Substituting the RR given by (\ref{RR-q}) in (\ref{virial-8}) we are led to
\ben \label{gov-a1} \frac{\partial \alpha}{\partial \lambda_k }\,=
\,-\langle {x}^{k}\rangle \hspace{0.5cm} \hspace{1.cm}\longrightarrow
\hspace{1.cm} \alpha \,=\,~\sum_{k=1}^{M}\,\left(1+\frac{k}{2}\right)
~\lambda_k ~\frac{\partial \alpha }{\partial \lambda_k} \,.\een  \nd
\nd and a complete solution is given by
\ben \label{gov-9a} {\alpha}(\lambda_1, ... , \lambda_M ) = ~
 \sum_{k=1}^{M}~D_k~{\left| \lambda_k \right|^{{2}/{(2+k)}}}~,\een
where the $D_k$s are positive real numbers (integration
constants). The $\alpha$-domain is ${\it D}_{\alpha}=
\left\{(\lambda_1,\cdots, \lambda_M ) / \lambda_k ~\in ~\Re
\right\}=\Re^M$. Also, Eq.(\ref{gov-9a}) states that for
$\lambda_k < 0$, $\alpha$ is a monotonically decreasing function
of the $ \lambda_k $, and as one expect from the Legendre
transform of $I$, we end up with a concave function. We may obtain
the $\langle {x}^{k}\rangle$'s from the reciprocity relations
(\ref{RR-q}). For $\lambda_k ~ < ~ 0 ~$ one gets \ben
\label{prop-2a-RR}
 \langle x^k \rangle~=~-~\frac{\partial \alpha}{\partial \lambda_k} ~=~
\frac{2}{(2+k)}~ D_k~ ~ \left| \lambda_k \right|^{-~k/(2+k)} ~
>~0. \een
\nd and then, using (\ref{eq.1-12}) one us able to build up  $I$.

\vskip 3mm \nd The general solution for $\alpha$ - PDE exists.
Uniqueness is, again, proved from an analysis of the associated
Cauchy problem \cite{ArX2Alpha}. Thus, Eq. (\ref{gov-9}) implies
once more a kind of ``universal" prescription, a linear PDE that
all SE-eigenvalues
 must necessarily comply with.

\vskip 3mm

\nd {\it The mathematical structure of the Legendre transform}
leads to a relation between the integration constants $C_k$ and
$D_k$ pertaining to the $I$ and $\alpha$ expressions,
respectively,  given by (\ref{gov-9}) and (\ref{gov-9a}). In
\cite{ArX2Alpha} we studied with some detail this relation. In our
two scenarios, $\left\{I, \langle {x}^{k}\rangle \right\}$ and
$\left\{\alpha, \lambda_k \right\}$, we have \cite{ArX2Alpha}
 \ben
\label{IX-AL-1} ~C_k~=~\frac{k}{2}~\bar{C}_k~,\hspace{1.5cm}
D_k=~\frac{k+2}{2}~\bar{D}_k~,\hspace{1.5cm} {\rm with}
\hspace{0.5cm} ~\bar{D}_k^{~(2+k)}~=~\bar{C}_k^{~k}~\equiv F_k^2.
\een \nd Consequently,  expressions (\ref{gov-9}) and
(\ref{gov-9a}) take the form, \ben \label{IX-AL-8} {I} ~=~
\sum_{k=1}^{M}~\frac{k}{2}
 ~{ \left[\frac{F_k}{\left| \langle {x}^{k}\rangle \right|} \right]^{2/k}}~,
 \hspace{1.5cm}
{\alpha} ~=~ \sum_{k=1}^{M}~\frac{k+2}{2}~
{\left[F_k~\left|\lambda_k \right|\right]^{{2}/{(2+k)}}}.\een \nd
The reciprocity relations (\ref{prop-2-RR}) and (\ref{prop-2a-RR})
can  thus be economically summarized in the fashion \ben
\label{IX-AL-10} F_k^{~2}~=~ \left|\lambda_k \right|^{k}~|\langle
{x}^{k}\rangle |^{(2+k)}~.\een

\section{4. Present results}

\subsection{The reference quantities $F_k$}

\nd The essential FIM feature is undoubtedly  its being an estimation measure known to obey the Cramer Rao (CR) bound of Eq. (\ref{rao})
 \cite{frieden2}. 
   Accordingly, since our partial differential equation has multiple solutions, it is natural to follow Jaynes's MaxEnt ideas and  select amongst them the one that optimizes the CR bound, that constitutes the informational operative constraint in Fisher's instance. Of course, Jaynes needs to maximize the entropy instead. We will also, without loss of generality,
   renormalize the reference quantities $F_k$. This procedure is convenient because it allows us to regard these quantities as statistical weights that
   optimize the CR-bound.
    In other words, our procedure entails that we extremize
 \ben \label{IX-AL-11} f(F_1,\cdots,F_M)=I~\left(\langle
{x}^{2}\rangle- \langle {x}\rangle^{2} \right) =
\sum_{k=1}^{M}~\frac{k}{2}
 ~{ \left[\frac{F_k}{\left|\langle {x}^{k}\rangle \right|} \right]^{2/k}}~
 \left(\langle {x}^{2}\rangle- \langle {x}\rangle^{2} \right)~.\een

 with the constraint
\ben \label{IX-AL-12}
\phi(F_1,\cdots,F_M)&=&\sum_{k=1}^{M}~F_k^{~2/k}=1.\een

\nd We are going to apply now the preceding considerations so as
to  obtain the eigenvalues of the quartic anharmonic oscillator.

\subsection{Quartic anharmonic oscillator}

\nd The Schr\"odinger equation for a particle of unit mass in a
quartic anharmonic potential reads, \be \label{AHO4-1}
\left[-~\frac{1}{2}~\frac{\partial^2~}{\partial x^2}
+~\frac{1}{2}~k~ x^2~+ \frac{1}{2}~\lambda ~x^4 \right]\psi~= ~ E~
\psi. \ee \nd where $\lambda$ is the anharmonicity constant.
According to \cite{ArX2Alpha,ArX3}, we can ascribe to
(\ref{AHO4-1}) a Fisher measure
 and make then the following identifications:
$~ \alpha = 8 E \,,~
  \lambda_2= -4~k \,,~ \lambda_4= -4~\lambda.$ Accordingly, we have, in the
$\left\{\alpha, \lambda_k \right\}$ - scenario [Cf.
(\ref{IX-AL-8})],  \ben  \label{AHO4-2} \alpha =
~2~F_2^{1/2}~\left|\lambda_2 \right|^{1/2}+3~F_4^{1/3}
~\left|\lambda_4 \right|^{1/3}. \een

\nd  The functions $f$ and $\phi$ defined by (\ref{IX-AL-11}) and
(\ref{IX-AL-12}), respectively, can here be recast [using
(\ref{IX-AL-10})]  as \ben \label{AHO4-3}
f(F_2,F_4)&=&F_2+2F_2^{1/2}F_4^{1/3}~\left|\lambda_2
\right|^{-1/2}\left|\lambda_4 \right|^{1/3} ~, \n \\
 \phi(F_2,F_4)&=&F_2+F_4^{1/2}~. \n \een

\nd After these preparatory moves we can recast our methodology
 in a convenient specialized fashion, suitable for the task at hand. We just face the
 simple two-equations system:
  \ben \label{AHO4-4} \left \{ \begin{array}{l}
\vec{\nabla}f(F_2,F_4)=\mu \vec{\nabla}\phi(F_2,F_4)\\
\hspace{0.2cm} \phi(F_2,F_4)~=~1\\ \end{array} \right. \een
 where $\vec{\nabla} \equiv \left( \partial_{F_2},\partial_{F_4} \right)$.
 Straightforward solution of it yields

  \ben \label{AHO4-5}F_2^{-1/2} (1-F_2)^{-1/3} \left(7F_2-3\right) =~3~\left|\lambda_2
\right|^{1/2}\left|\lambda_4 \right|^{-1/3}~,\hspace{1.cm}
F_4=(1-F_2)^2~,\hspace{0.5cm}\een from which we obtain $F_2$ and
$F_4$. Substituting them into (\ref{AHO4-2}) we determine $\alpha$
and, of course, the eigenvalue $E=\alpha/8$. Consider now our SE
(\ref{AHO4-1}), taking $k=1$ and a given value of $\lambda$
($0.0001 \leq \lambda \leq 10000$). The function [Cf. Eq.
\ref{IX-AL-11}] $f(F_2,F_4)=I~\langle {x}^{2}\rangle$ exhibits, as
a function of its arguments, a unique ``critical" point that
satisfies (\ref{AHO4-5}). Using $f=f_{critical}$, that optimizes
the CR-bound, we find a ground-state eigenvalue that is in good
agreement the literature. In this way, after properly dealing with
(\ref{IX-AL-8}), with the $F_k$ regarded as ``FIM statistical
weights" that optimize the Cramer Rao inequalities, we determine
$\alpha$ as a function of the $\lambda_k$ {\it without passing
first through a Schr\"odinger equation}, which is a notable
aspect of the present approach.

\nd Interestingly enough, the Cramer-Rao inequality us equivalent
to the quantum uncertainty principle (see the Appendix for details
and references). Thus, our methodology actually employs
Heisenberg's celebrated principle to pick up just one solution
among the several ones that our partial differential equation
possesses.

 \vspace{0.7cm}

\nd \begin{minipage}{9.60cm}
\small
\nd  {\bf Table}: Ground-state eigenvalues of the SE
(\ref{AHO4-1}) for $k=1$ and several
values of the anharmonicity constant $\lambda$. The values of the second column correspond to those one
finds in the literature,  obtained via a numerical approach to
the SE. {\it These} results, in turn,  are
nicely reproduced by some interesting theoretical approaches that,
however, need to introduce and adjust some empirical constants
\cite{Banerjee}. Our values, in the third column, are obtained by
means the present theoretical, parameter-free procedure. The
fourth column displays  the associated Cramer-Rao bound.
\rm
\end{minipage}
\begin{minipage}[t]{6.70cm}
\footnotesize
\begin{flushright}
\begin{tabular}{||l||c||c|c||}
\hline
\hline
\hspace{0.4cm}$~\lambda~$ &~$~E_{num}~$~ & $E=\alpha /8$ &
$f=I~\langle {x}^{2}\rangle$ \\
\hline
\hline
0.0001&~1.000074&1.000074&1.000059\\
0.001&~1.000748&1.000739&1.000591\\
0.01&~1.007373&1.007263&1.005824\\
0.1&~1.065285&1.063047&1.051255\\
1&~1.392351&1.353533&1.296590\\
10&~2.449174&2.213973&2.040974\\
100&~4.999417&4.212932&3.782394\\
1000&10.639788&8.587748&7.599439\\
\hline
\hline
\end{tabular}
\end{flushright}
\rm

\end{minipage}

\vspace{0.5cm}

\section{5. Conclusions}

\nd On the basis of a variational principle based on Fisher's
information measure, free of adjustable parameters, we have
obtained the Schr\"odinger energy-eigenvalues for the fundamental
state of the quartic anharmonic oscillator (for several
anharmoniticy-values). Our theoretical results, obtained without
passing first through a Schr\"odinger equation, are in a good
agreement with those of the literature.
 This constitutes an illustration of the power
of information-related tools in analyzing physical problems.

\nd Thus, we have in this communication introduced a new general
technique for eigenvalue-problems of linear operators, whose use
seems  to constitute a promising venue, given the results here
displayed.

\vspace{0.5cm}


\nd {\bf Acknowledgments-} This work was partially supported by the
Projects FQM-2445 and FQM-207 of the Junta de Andalucia
(Spain, EU).

\vspace{0.5cm}


\section{Appendix: Cramer-Rao and Uncertainty Principle}

\nd It is well known that the Cramer-Rao inequality may be
regarded as an expression of Heisenberg's Uncertainty Principle
(See, for instance,  \cite{frieden2}). Remember that a precise
statement of the position-momentum uncertainty principe reads
\cite{Mathews} \ben \label{CR-HP-1} (\Delta x)(\Delta p) \geq
\frac{\hbar}{2}\hspace{1.3cm}or \hspace{1.3cm} (\Delta x)^2
(\Delta p)^2 \geq \frac{\hbar^2}{4}, \een where
\ben \label{CR-HP-2}(\Delta x)^2
&=&\left\langle \left( x-\langle x\rangle \right)^2 \right\rangle =
\langle x^2\rangle-\langle x\rangle^2 \\
\label{CR-HP-3} (\Delta p)^2
&=&\left\langle \left( p-\langle p\rangle \right)^2 \right\rangle =
\langle p^2\rangle-\langle p\rangle^2. \een

\nd In a one-dimensional configuration-space, if $\psi$ is a
normalizable real wave function,  \ben \label{CR-HP-4} \langle
p\rangle &=& \langle - i \hbar \frac{\partial ~}{\partial
x}\rangle
= - i \hbar \int \psi \frac{\partial ~}{\partial x} \psi ~ dx
= -i \frac{\hbar}{2} \int  \frac{\partial ~}{\partial x} \psi^2 ~ dx = 0~,\\
\label{CR-HP-5} \langle p^2\rangle &=& \left\langle - \hbar^2
\frac{\partial^2 ~}{\partial x^2}\right\rangle = - \hbar^2
\int{\psi \frac{\partial^2 ~}{\partial x^2} \psi~dx}~. \een
Substituting (\ref{CR-HP-4}) and (\ref{CR-HP-5}) in
(\ref{CR-HP-3}) and using  (\ref{eq.1-7}) leads to the above
mentioned  connection between the uncertainty in momentum $\Delta
p$ and the Fisher's measure $I$, i.e., \ben \label{CR-HP-6} \left(
\Delta p \right)^2 =  - \hbar^2 \int  \psi \frac{\partial^2
~}{\partial x^2} \psi~dx= \frac{\hbar^2}{4}~I ~.\een \nd If  this
 relation is substituted into (\ref{CR-HP-1}) we immediately
 arrive to the the CR-bound,
\ben  (\Delta x)^2 (\Delta p)^2 \geq \frac{\hbar^2}{4}
\hspace{1.3cm} \longrightarrow \hspace{1.3cm}
I (\Delta x)^2 \geq 1.  \een

\nd Coming now back to the $\{ \alpha,\lambda_k \}$-scenario, one
easily ascertains that
 Eq. (\ref{IX-AL-11}) can be given a clear ``Heisenberg's aspect"
\ben \label{goetingen} f(F_1,\cdots,F_M)=
\sum_{k=1}^{M}~\frac{k}{2}
~{ \left[~{F_k}~{\left| \lambda_k \right|} ~\right]^{2/(2+k)}}
\left(F_2^{~1/2} {\left| \lambda_2 \right|^{-1/2}} -
F_1^{~4/3} {\left| \lambda_1 \right|^{-2/3}} \right)~.\n \een

\end{document}